# Study of Beam Loss Monitors (BLM) in Storage Ring

**S. M. Esmaeili[c], S. A. H. Feghhi[c]**

[c]*Radiation Applications Department, Shahid Beheshti University,Evin, G. C. Tehran, Iran*

ABSTRACT: The Beam Loss Monitors (BLM) are designed to measure the position and amount of beam loss in accelerators. In this article, we have studied the 3 GeV electron losses in the storage ring and secondary particles from the losses on the beam pipe. We have compared ionization chamber, NaI and Si radiation detectors as BLM and selected Si detector for further studies. We have calculated electron deflection angle due to magnetic field mismatches in dipole magnets, quadrupoles and sextupoles and assumed that electron beam is deflected and hit the beam pipe with the angle of 3 degrees with respect to the beam axis. The number and energy of photons and secondary particles on beam pipe and in Si detector are calculated by the MCNP code and reported in this paper.

## 1. Introduction

With the information provided by the BLM placed around the storage ring, beam loss distribution pattern can be directly monitored [1]. We can use beam loss distribution pattern to improve the performance of storage ring in terms of life time and stability [1]. Beam losses can be caused by collision of the beam to the diagnostics equipment such as Faraday cups, wire scanners, scrapers or any other devices which are located on the beam path or by other reasons such as residual gas scattering, misalignment, instabilities, and Halo scraping [2][3][4][5].
In this paper, we have studied the losses due to the magnet misalignment, magnet vibration and magnet current fluctuations that causes the beam to experience different magnetic fields [6].
We have calculated the deflection angle of electron beam due to the magnetic field mismatches in dipole, quadrupoles and sextuples magnets geometrically to be around 0.5 to 6 degrees. In the following calculations we have considered the electron typical deflection angle of 3 degrees.
We have calculated the number and energy of secondary particles generated from hitting of the deflected 3GeV electrons to the storage ring beam pipe via MCNP code [7]. In all calculations typical storage ring's parameters are considered [8][9][10][11]. According to the simulation results, 34.8 electrons, 139.7 photons and 0.1 neutrons are exited from the beam pipe with the total energy of 1073.3 MeV, 1467.9 MeV and 1.3 MeV, respectively.

To detect such losses, beam position monitors are one of the essential sensors, but their information is not sufficient, that is why we need other types of radiation detectors around the storage ring [12][13][14][15][16]. Radiation detectors such as scintillators or silicon detectors are found in variety of applications such astronomy, nuclear physics, medicine and even air pollutions [17][18][19][20][21][22].
In the following we study ionization chamber, NaI and Si detectors as beam loss monitors. Among them, Si detector is selected in our studies due to its advantages of high resolution, high efficiency, linear response in a wide range of energy, relatively fast response, being not sensitive to the magnetic field, and possibility to construct in various forms [23]. The Si detector is placed 10 cm away from the beam pipe and the number and energy of photons and secondary particles that would reach the detector are calculated by MCNP code [7]. It is concluded that only 0.05 electrons, 0.9 photons and 0.57 neutrons with the total deposited energy of 0.16 MeV, 3.89 MeV and 0.003 MeV respectively will reach the detector.

## 2. Loss Angle

The angle of loss is the angle with respect to the beam axis at which the electron particle is separated from the beam and collides with the accelerator beam pipe. This angle is essential for calculation of the secondary particle characteristics. Beam loss may occur as a result of magnet vibrations, electrical current fluctuations in magnets winding or magnet misalignment [24]. These factors make electrons to experience a different magnetic field than the optimal and normal value when they pass through magnets. So, if the magnetic field applied to the particles are less or more than nominal, particles do not travel their pre-defined path, and may hit the beam tube and become lost. Therefore, by changing the values of magnetic field, the loss angle could be obtained.

## 2.1. Loss angle due to dipole magnet mismatch

In order to obtain the loss angle and the collision point of electron on the beam tube, it is assumed that the dipole magnet's magnetic field to be less or greater than its nominal value of 0.748 T in the storage ring [8]. By changing the magnetic field value, electrons travel circular paths with smaller or larger radii and collide to the beam pipe and become lost.
In the our storage ring study case, the bending angle of dipoles is 3.6 degrees, therefore, the total number of 100 dipole magnets will be installed around the storage ring [8]. Figure 1 shows the schematic placement of dipole and quadrupole magnets with the distance of 30 cm from each other in storage ring lattice [8]. In order to study the beam loss due to magnetic field mismatch in dipole magnets, three scenarios can be followed: particles pass through the dipole magnet and enter the quadrupole without any loss (See Fig. 1), particles collide the beam tube before entering the quadrupole (See Fig. 2), and particles collide the beam tube inside the dipole magnet (See Fig. 3).

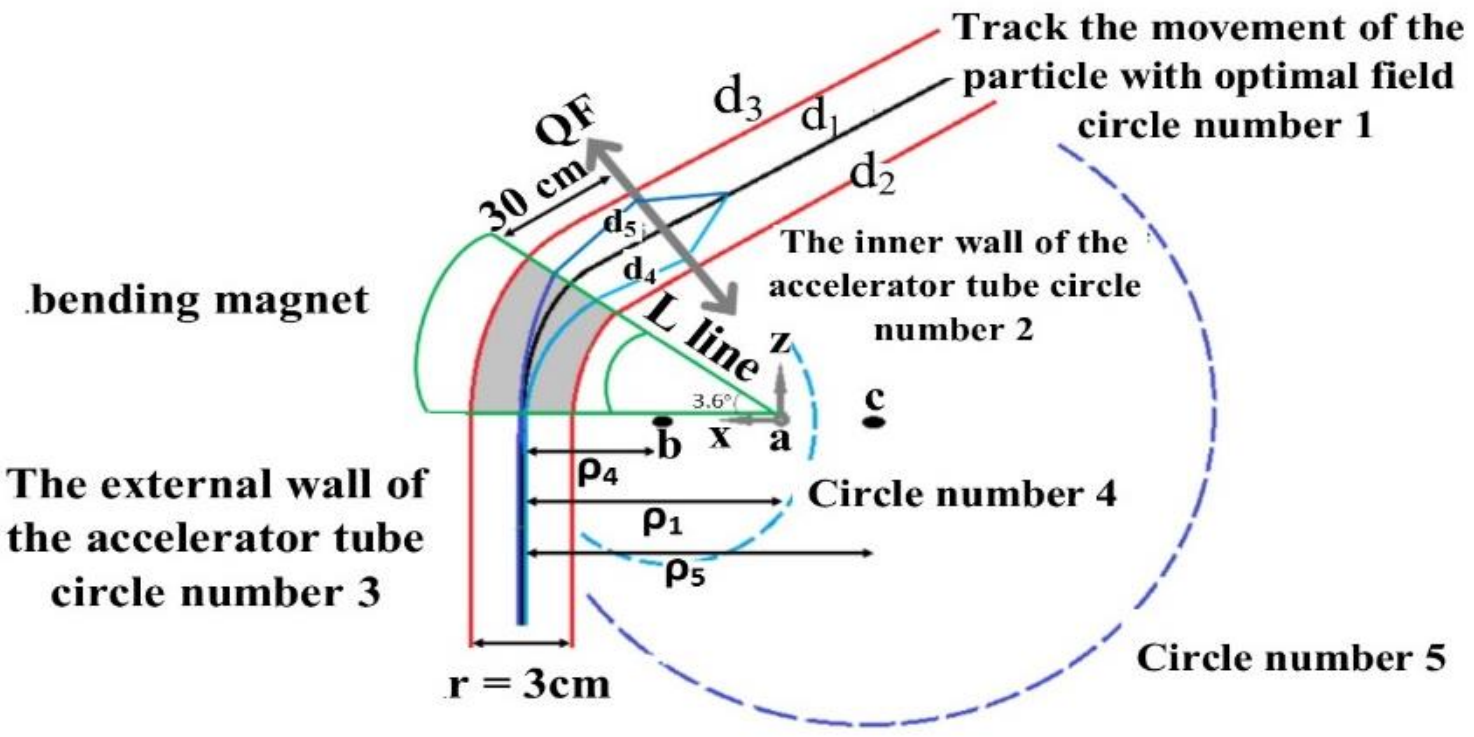

**Figure 1.** The schematic placement of dipole and quadrupole magnets. In this figure, particles enter to the quadrupole without loss.

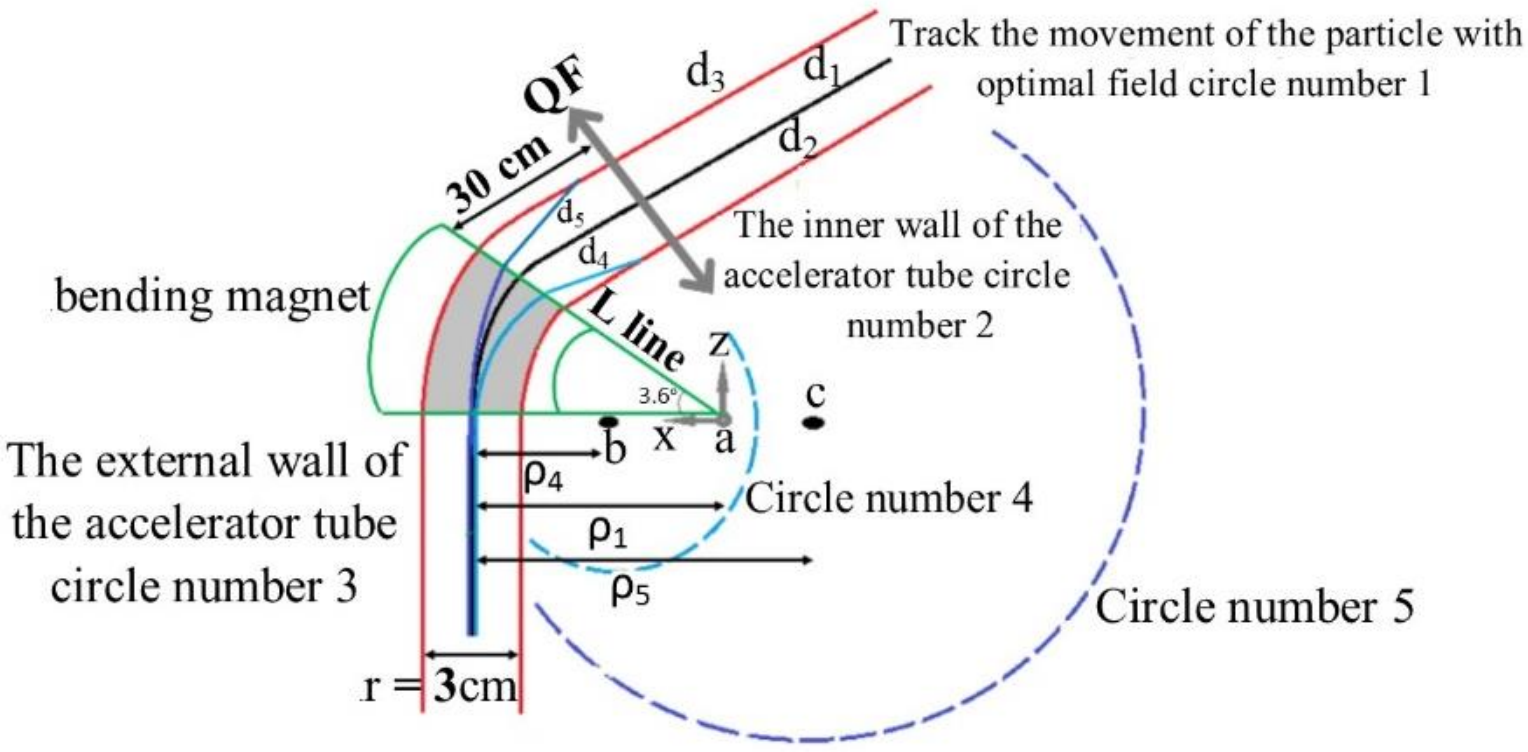

**Figure 2.** Particles collide to the beam tube after dipole magnet and before entering the quadrupole.

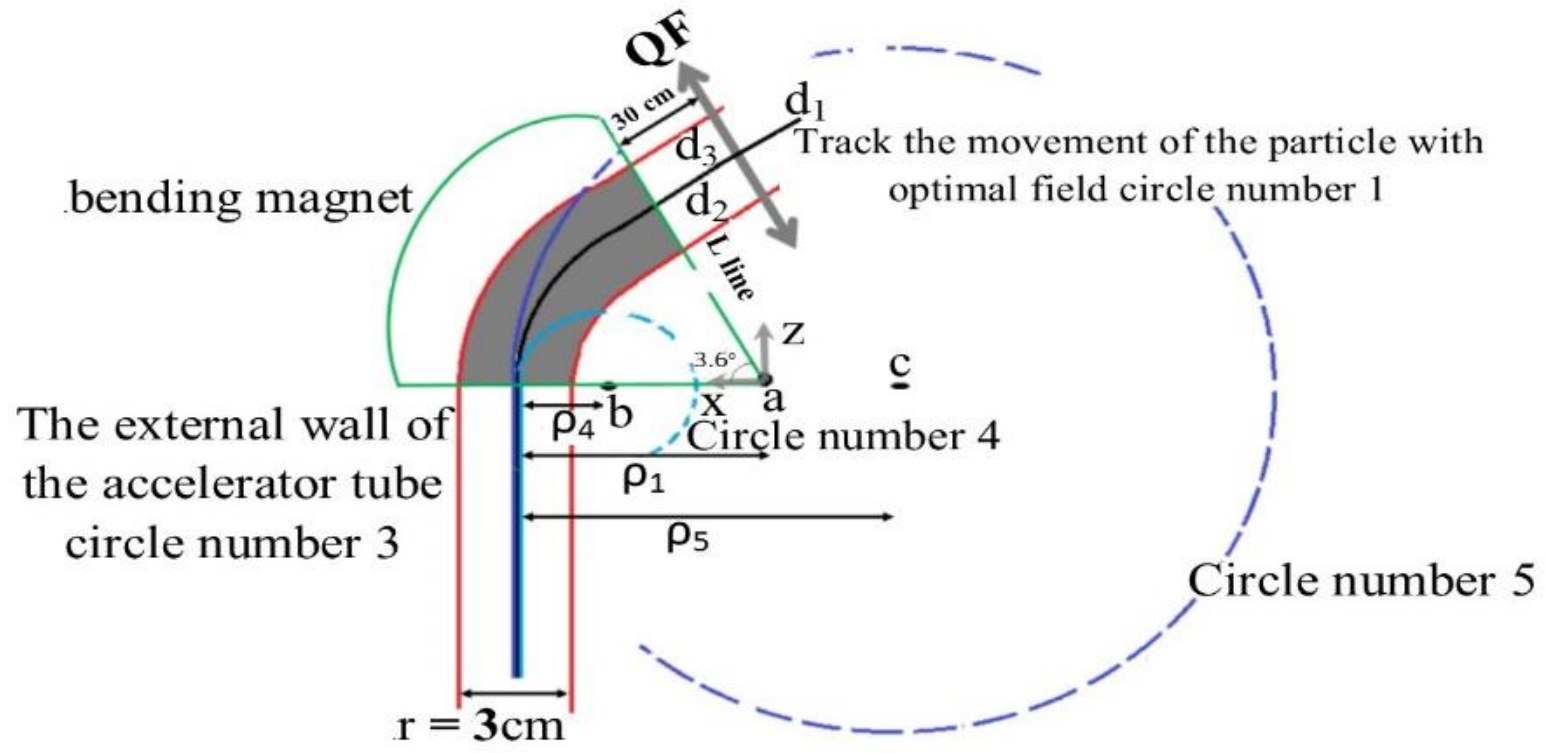

**Figure 3.** Particles collide to the beam tube inside of dipole magnet.

In these figures, the solid line at the center of beam pipe represents a reference path that particles should pass through without any disturbance. Inside the dipole magnet particles pass on a circular path (with the radius of $\rho_1$) and outside of dipoles, they pass on straight lines ($d_1$). The radius of circular path can be obtained by $\rho = \frac{\gamma m_0 C}{eB}$ where $\gamma$ is Lorentz factor and $m_0$ is electron rest mass energy (0.511 MeV), C is speed of light, e is electron charge and B is dipole magnetic field [24]. However, the calculations for superconductive types can be different [25][26][27]. The inner and outer shells of beam pipe inside the dipole magnet considered as circles with the radiuses of $\rho_2$ and $\rho_3$, respectively, and outside the dipole magnets, they are considered as straight lines $d_2$ and $d_3$. If the magnetic field of dipole magnet is larger than the nominal value, particles will become lost because of hitting the inner shell of the beam pipe and will pass the circular path with the radius of $\rho_4$ inside the magnet and the straight line of $d_4$ outside and if it is lower than the nominal value, particles will hit the inner shell of the beam pipe and pass the circular path with the radius of $\rho_5$ inside the magnet and straight line $d_5$ outside. Point "a" is considered as our geometrical reference point which is the center of circle "1" (See Fig. 1, 2, 3). Line "L" is considered as the border of dipole magnet where particles travel on a straight line after that. The diameter of beam pipe is considered to be of 3 cm [28][29].

First, we have calculated the equation of circles "1", "2", "3" and lines "$d_1$", "$d_2$", "$d_3$" and "L". Then for each case of magnetic field, we obtain the equation of circles 4 and 5 in X-Z plane. For a nominal magnetic field of 0.748 T inside of dipole magnets, $\rho_1, \rho_2, \rho_3$ are 1338.838 cm, 1337.338 cm and 1340.338 cm respectively. Equation of line "L" is obtained to be x=15.894z. The z coordinate of intersections between line "L" and circles "2" and "3" are 83.975 cm and 84.163 cm, respectively. Then by calculating the intersection points of line "L" with particle paths (circles 4 and 5) and obtaining their z coordinates, we can conclude that the particles will pass through the dipole magnets if 83.975 cm<z<84.163 cm, otherwise they are lost inside the magnet (See Fig. 3).

For these purposes, we have considered the dipole magnets to have magnetic field values between 0.13 T and 1.4 T. Based on our studies, for 0.7481 T ≤ B ≤ 0.99 T and 0.51T ≤ B ≤ 0.7479, particles will enter the quadrupoles without any loss. For 1 T ≤ B ≤ 1.19 T and 0.33 T ≤ B ≤ 0.5 T, particles will hit the inner and outer shells of beam pipe at the distance between dipole and quadrupole magnet. If 1.2 T ≤ B or B ≤ 0.32 T, particles will hit the beam pipe inside the dipole magnet. Table I shows the calculated deflection angle of electrons for different dipole magnetic field values.

**TABLE I.** deflection angle of electrons with respect to the reference axis for different dipole magnetic fields

| Magnetic field (T) | Deflection angle (Degrees) |
|---|---|
| 1 | 1.212 |
| 0.33 | 5.189 |
| 1.19 | 2.174 |
| 0.5 | 6.007 |
| 1.2 | 2.108 |
| 0.32 | 2.05 |

## 2.2. Loss angle due to quadrupole and sextupole magnets mismatch

The main task of quadrupole magnets is the convergence or divergence of the beam. Defocusing (QD) and focusing (QF) magnets, converge particle beams in one direction (z axis for QD and x axis for QF) and diverge them in the other (z axis for QF and x axis for QD) [24]. Due to energy dispersion, particles with different energies passing through quadrupoles will deflect with different angles. Higher energy particles are more rigid than lower energy one for the same magnetic field value. That is why sextupoles are used where particles with higher energy and lower energy experience higher and lower magnetic fields respectively that makes the convergence procedure more efficient [30][31].
Figure 4 shows the schematic placement of quadrupole and sextupole magnets in storage ring [8]. Defocusing magnet (QD) converge particles in the vertical direction (z) and diverge in the horizontal direction (x). The probability of beam loss in the distance between defocusing magnet and sextupole is ignorable and the beam loss after passing the focusing quadrupole is also impossible since it will converge particles in the horizontal direction (x). Then if there is a loss in horizontal direction (x), it would be in the distance between sextupole and focusing quadrupole.

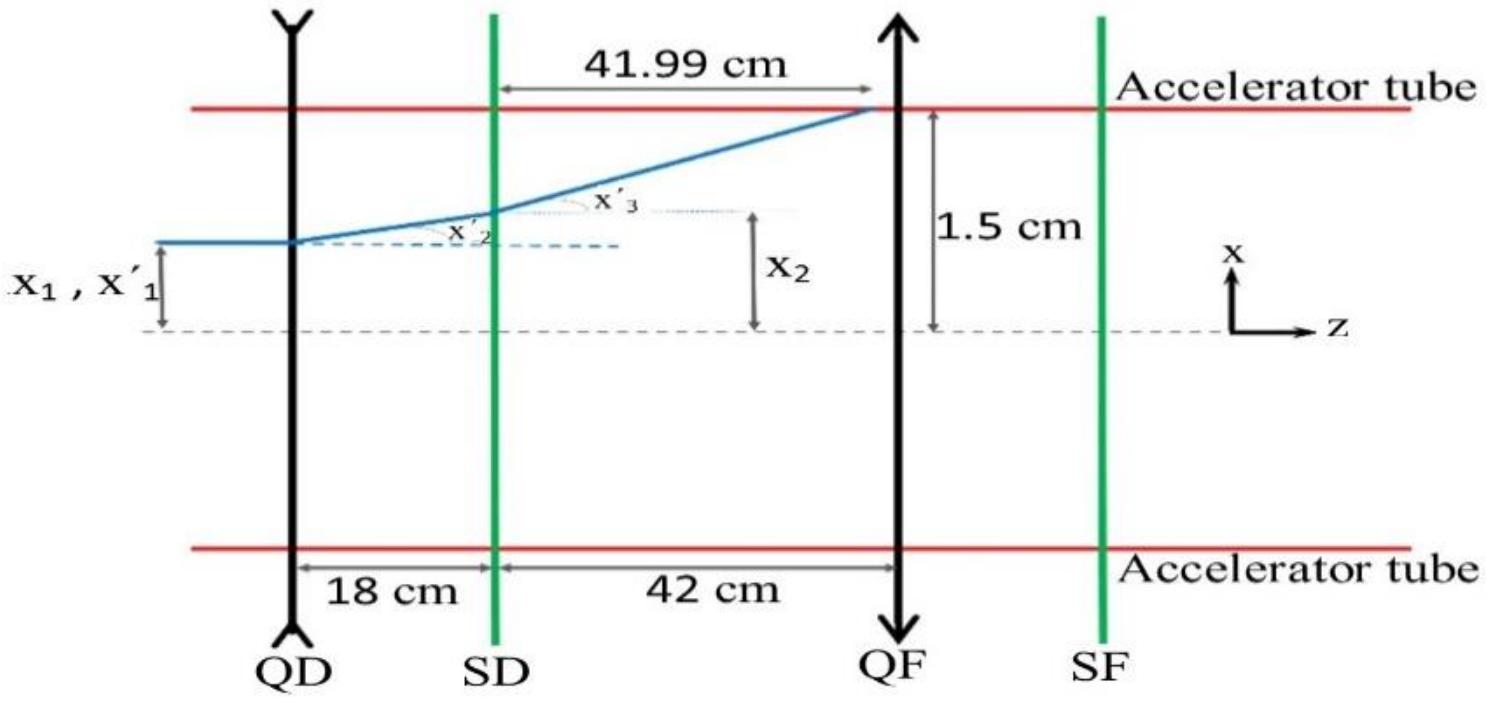

**Figure 4.** Schematic placement of quadrupole and sextupole magnets in storage ring.

We have calculated that if particle collide the beam pipe 0.01 cm before the focusing quadrupole (41.99 cm after the sextupole-See Fig. 4). It is due to the magnetic field of 1.248 T inside the sextupole. Then based on $B_x=K_s.z^2$, $B_z=K_s.X^2$, and $K_s$=8.352 T/cm$^2$, particle should be at the distance of 3.864 cm from the sextupole axis to experience the 1.248 T [24]. Table II shows the particle deflection angle according to its distance to the sextupole axis and consequent magnetic field of sextupole that particle passes from.

**TABLE II.** Loss angle and distance between particle collision point and sextupole location

| Magnetic field (T) | Distance from sextupole magnet axis (cm) | Loss angle | Distance between particle collision point and sextupole location (cm) |
|---|---|---|---|
| 1.248 | 0.3866 | 1.518 | 41.99 |
| 1.607 | 0.4386 | 1.599 | 38 |
| 2.152 | 0.5076 | 1.671 | 34 |
| 3.031 | 0.6024 | 1.713 | 30 |
| 4.098 | 0.7005 | 1.732 | 27 |

According to the calculations, as can be seen in Table I and II, the deflection angle varies based on the magnetic field from 0.5 to 6 degrees. Angle of deflected electron with respect to the beam line is one of the main parameters in calculation of secondary particles. Therefore, for calculation of the number and energy of secondary particles, we assume that the electron beam is deflected and hit the beam pipe by an angle of 3 degrees.

## 3. Secondary Particles

We used MCNP code [7] to obtain secondary particle and photon spectrums generated in the external beam pipe wall due to hitting of the deflected electrons. For this purpose, a hollow cylinder with the inner and outer radius of 1.5 cm and 1.7 cm is considered the beam pipe. The length of this cylinder is 4 m, and it is made of stainless steel 316. An electron spot source with an energy of 3 GeV and deflection angle of 3 degrees is considered at the center of this cylinder. Figure 5 shows the geometrical parts of our simulation in MCNP [7].

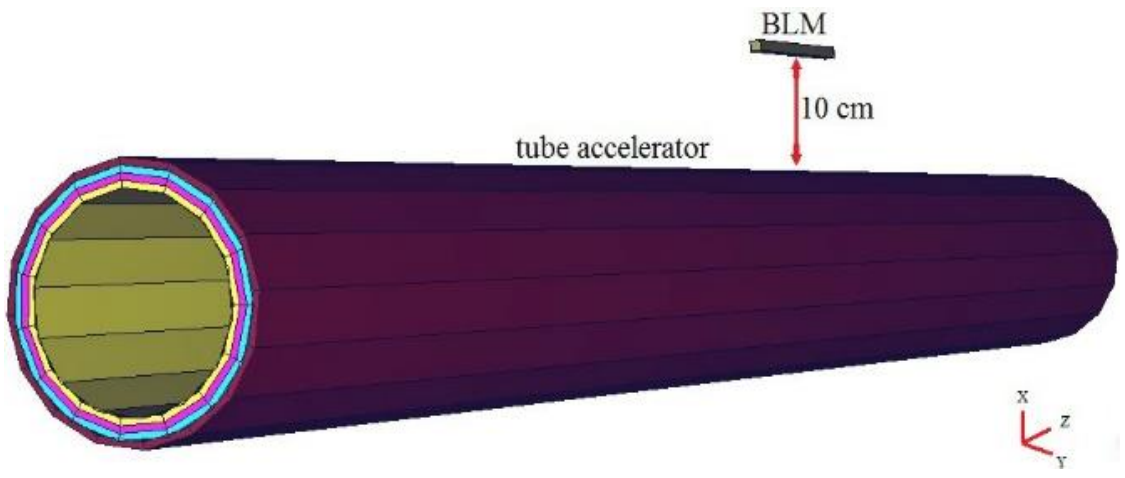

**Figure 5.** Simulated geometry of the accelerator tube and the BLM detector.

To analyze the secondary particles, as shown in Fig. 5, the 2 mm thickness of steel beam pipe is divided into equal quadrants of 0.5 mm thickness. As electron penetrates in the steel, it deposits all its energy in first quadrant and makes secondary particles. Due to 3 degrees deflection angle, electron passes a distance of 9.56 mm in the first quadrant. In this quadrant, the 1584.03 MeV of deposited electron energy converts to the photons and secondary electrons where they also stop in this quadrant and produce more secondary electrons and neutrons, and remaining amount of deposited energy, 1415.97 MeV, converts to the secondary electrons which exit from this quadrant. Deposited energy of electrons and photons in the first part is 94.4336 MeV, while this value is only 0.00349 MeV for neutrons. Therefore, the total deposited energy in the first 0.5 mm thickness of steel is about 94.43709 MeV.

In the second quadrant, from each 10000 entered secondary electrons, 9993 electrons deposit their whole energy, so that the amount of 756.998 MeV is converted to secondary electrons and photons. From each 100000 entered photons from the first quadrant, 65 photons deposit their whole energy inside second the quadrant, so that the amount of 120.387 MeV is deposited in this part. From each 10000 entered neutrons from first quadrant, 9986 neutrons deposit their whole energy in the second quadrant, so that the amount of 0.0025 MeV is deposited in this layer. As a result, the total energy of 120.3895 MeV is deposited in the second quadrant.
In the third layer, from each 10000 entered secondary electrons, 9910 electrons deposit their whole energy, so that the amount of 333.084 MeV is converted to secondary electrons and photons. From each 100000 entered photons from the previous part, 897 photons deposit their whole energy inside the third layer, so that the total amount of 127.289 MeV is deposited in this layer. From each 10000 entered neutrons from the second quadrant, 9990 neutrons deposit their whole energy in the third layer, so that the amount of 0.0033 MeV is deposited in this part. As a result, the total energy of 127.2923 MeV is deposited in the third quadrant.
In last layer, from each 10000 entered secondary electrons, 9659 electrons deposit their whole energy, so that the amount of 144.385 MeV is converted to secondary electrons and photons. From each 100000 entered photons from previous part, 3405 photons deposit their whole energy inside this layer, so that the amount of 105.610 MeV is deposited in this layer. From each 10000 entered neutrons from the third quadrant, 9992 neutrons deposit their whole energy, so that the amount of 0.0025 MeV is deposited in this part. As a result, total energy of 105.6125 MeV is deposited in the fourth layer of steel beam pipe.
The number and energy of deposited electrons, photons and neutrons in each quadrant of steel are given in Table. III and Table. IV. In order to calculate the energy fractions that electrons convert to photons and secondary electrons, in Table. III, only electron transportation is considered in the MCNP simulation.

**TABLE III.** The number and energy of deposited electrons for each tube layer in the case that only electron transportation is considered in the MCNP simulation.

| Tube layers | The number of deposited electrons | Deposited energy by electrons (MeV) |
|---|---|---|
| First 0.5 mm | 1 | 1584.03 |
| Second 0.5 mm | 0.99934 | 756.998 |
| Third 0.5 mm | 0.99102 | 333.084 |
| Fourth 0.5 mm | 0.96594 | 144.385 |

**TABLE IV.** The number and energy of deposited photons and neutrons for each tube layer.

| Tube layers | The number of deposited photons | The number of deposited neutrons | Deposited energy by neutrons (MeV) | Total deposited energy (MeV) |
|---|---|---|---|---|
| First 0.5 mm | 0 | 0.9966 | 0.0034 | 94.4370 |
| Second 0.5 mm | 0.00065 | 0.9986 | 0.0025 | 120.3895 |
| Third 0.5 mm | 0.00897 | 0.9990 | 0.0033 | 127.2923 |
| Fourth 0.5 mm | 0.03405 | 0.9992 | 0.0025 | 105.6125 |

Then according to calculations, finally from the outer surface of the beam tube, 34.8 electrons with a total energy of 1073.3020 MeV, 139.7 photons with a total energy of 1467.999 MeV and 0.1 neutrons with a total energy of 1.32 MeV may exit.
Figures. 6 and 7 show the spectroscopy of escaping electrons and photons from the accelerator tube in the energy ranges of 0-3000 MeV and 0-50 MeV, respectively.

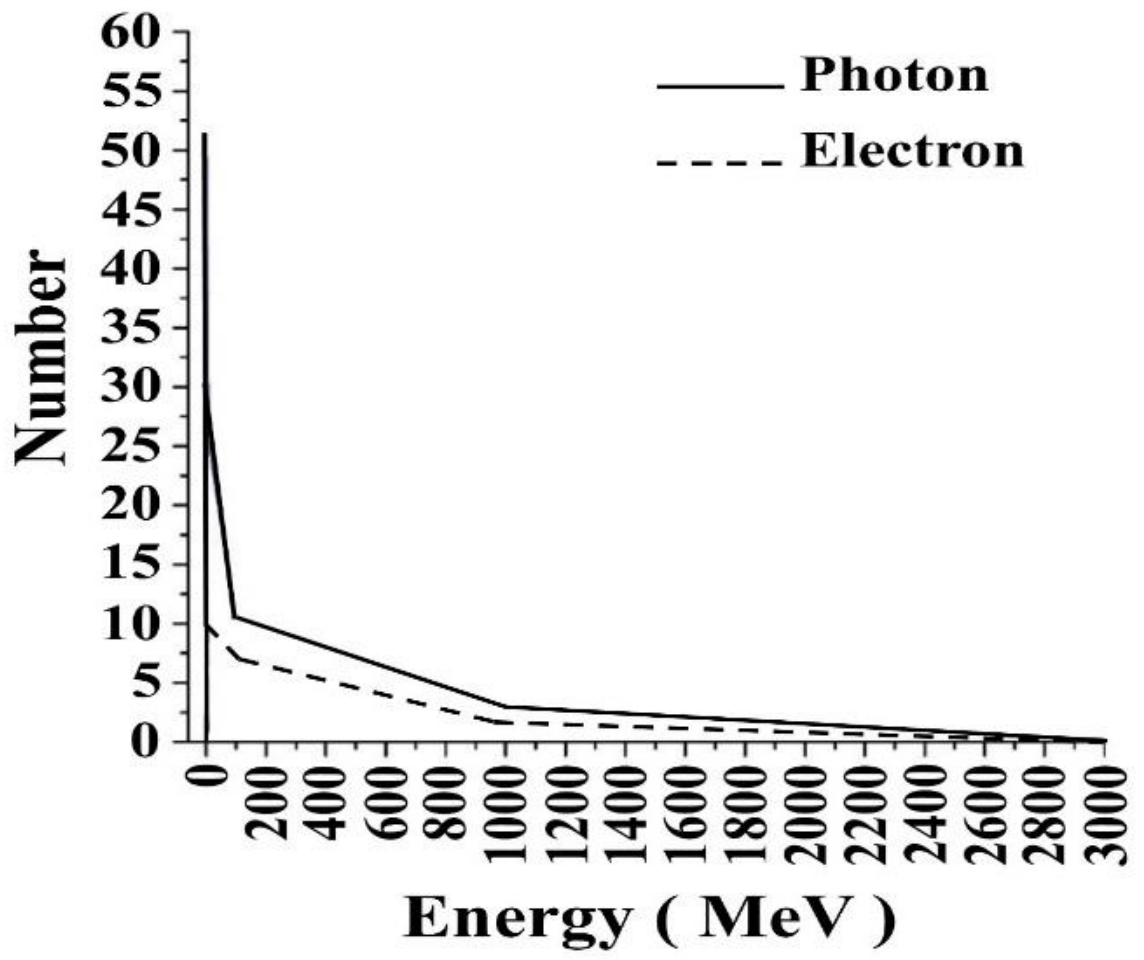

**Figure 6.** The spectroscopy of departing electrons and photons from accelerator tube for energies in the range of 0-3000 MeV.

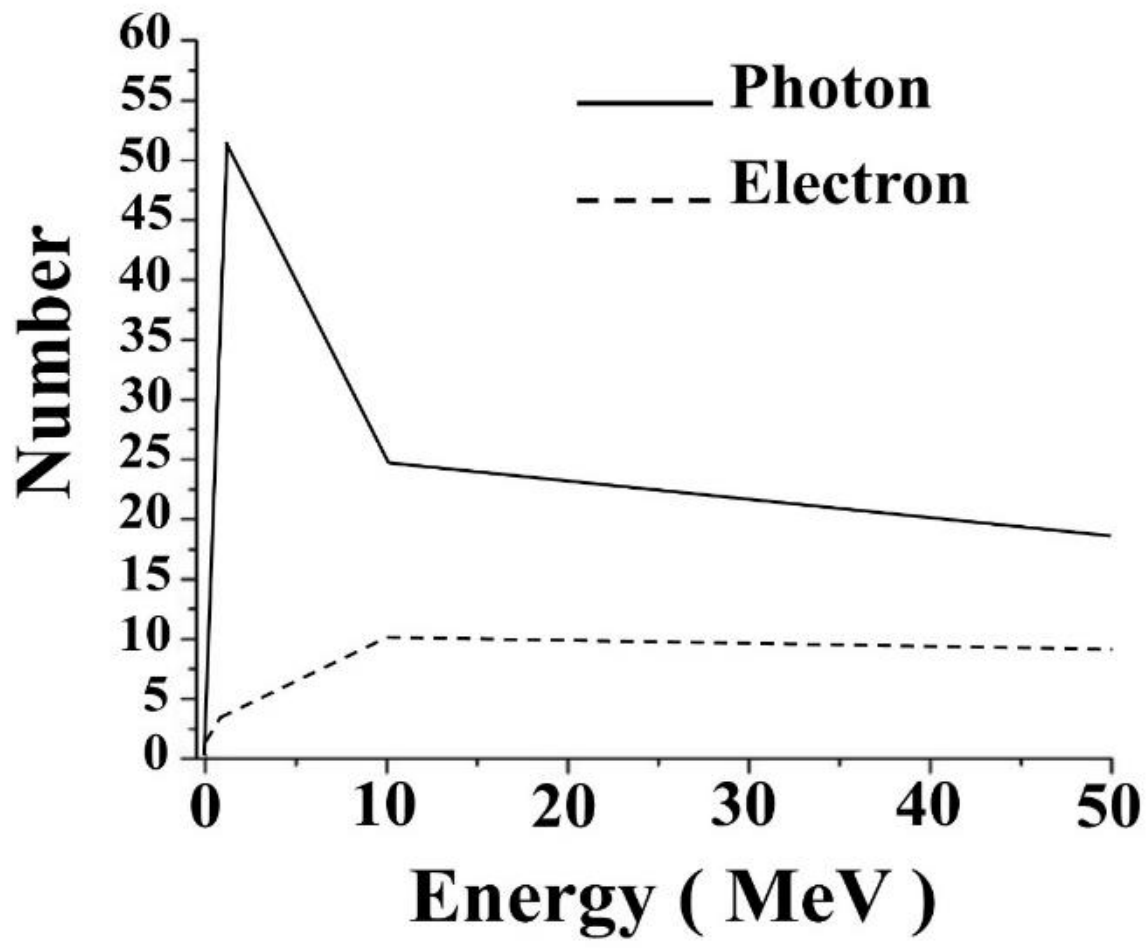

**Figure 7.** The spectroscopy of departing electrons and photons from accelerator tube for energies in the range of 0-50 MeV.

As shown in this figure, a large fraction of electron and photon particles escaping from the accelerator beam tube are in the energy range of 0-4 MeV. The spectroscopy of escaping neutron particles from accelerator tube is shown in Fig. 8. From this figure, it can be seen that the most exiting neutrons have an energy in the range of 0-100 MeV.

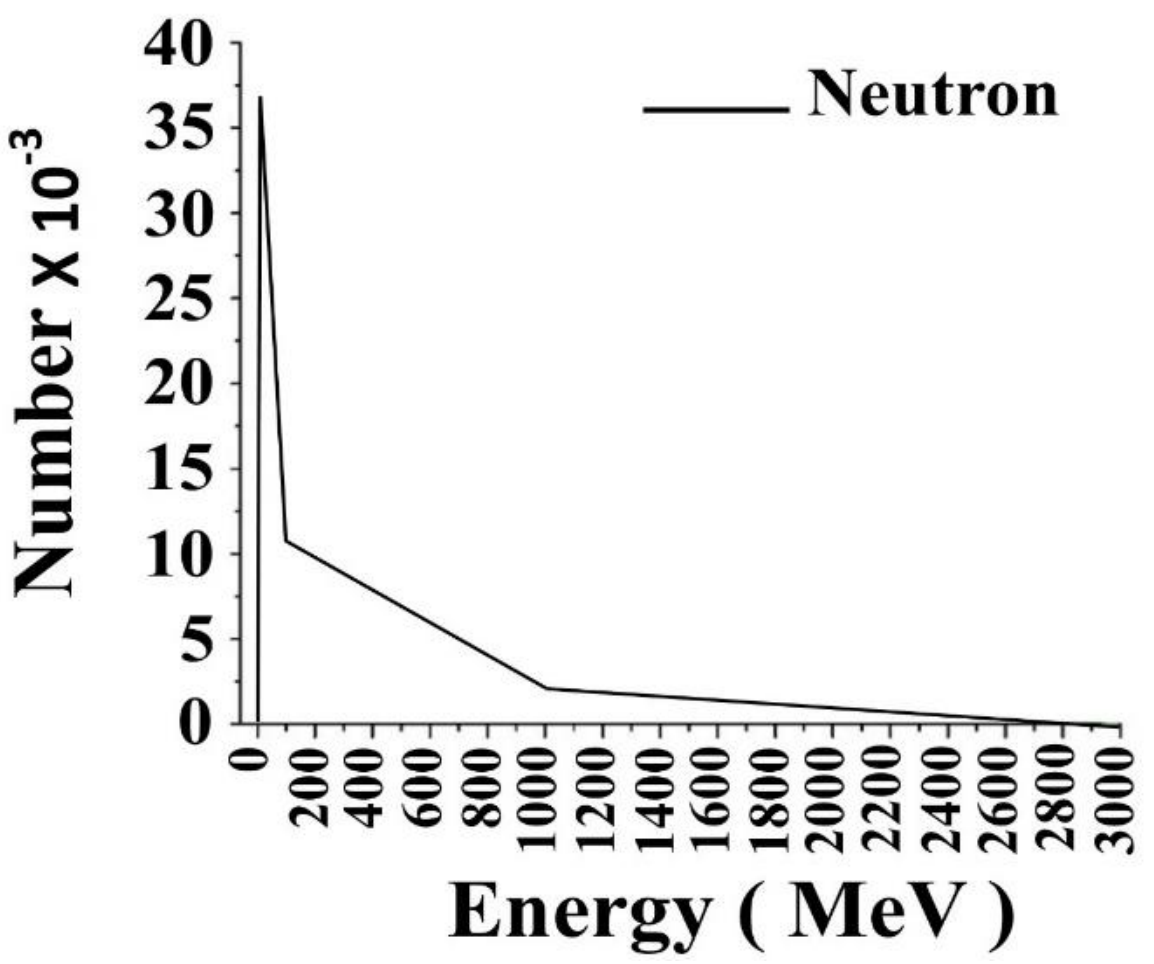

**Figure 8.** The spectroscopy of neutrons exited from outer shell of beam tube.

## 4. Particle Detectors as BLM

Among different particle detectors, ionization chamber, NaI and Si detectors have been compared in terms of the number and deposited energy of particles inside them. The number and deposited energy of the particles inside each detector have been investigated by the MCNP code [7]. For this purpose, three simulations were performed. In each of these simulations, the same geometrical and parameters described in section 3 are used. Detectors are placed on the four sides of the beam pipe (top, bottom, left and right), at a distance of 10 cm from the outer wall. In each simulation, all four detectors are considered to be of only one type (ionization chamber, NaI, or Si). Detectors are considered as cubes with the dimensions of 2 cm along X, 10 cm along Y and 10 cm along Z axis. In the space between two consecutive detectors, 2 cm of air is considered. The air is also located between the outer wall of the beam pipe and the detectors. Figure 9 shows the simulated geometry along with the labeled number of cells defined in the MCNP environment.

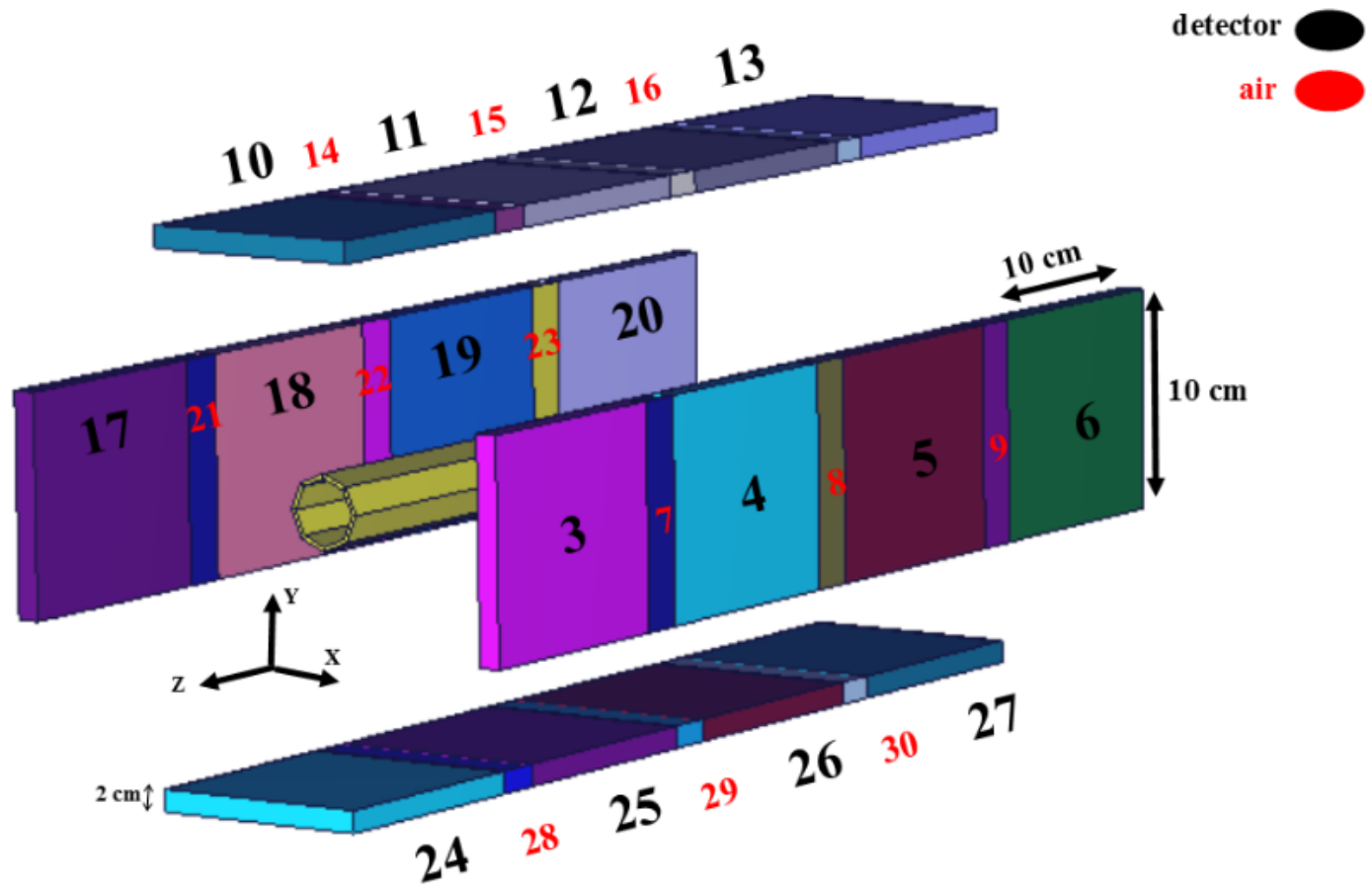

**Figure 9.** Simulated geometry with the labeled number of cells defined in the MCNP environment.

The 3 GeV electron beam passes on the Z axis inside the accelerating pipe and then will hit the beam pipe by the angle of 3 degrees and the secondary electrons and photons with their energy will be recorded inside the detectors. Figures. 10, 11, 12, 13 show the results of these simulations.

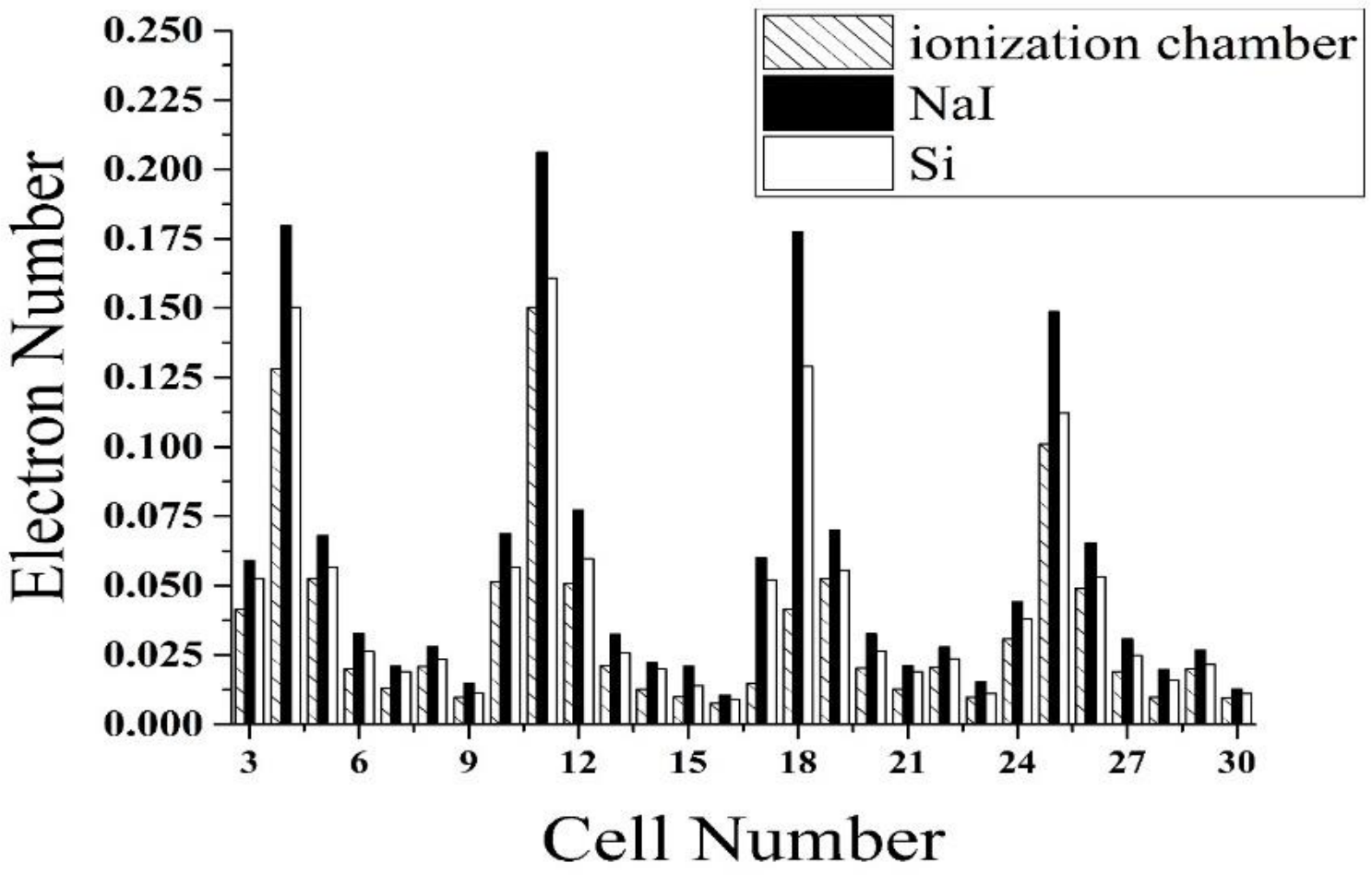

**Figure 10.** The number of deposited secondary electrons in detectors.

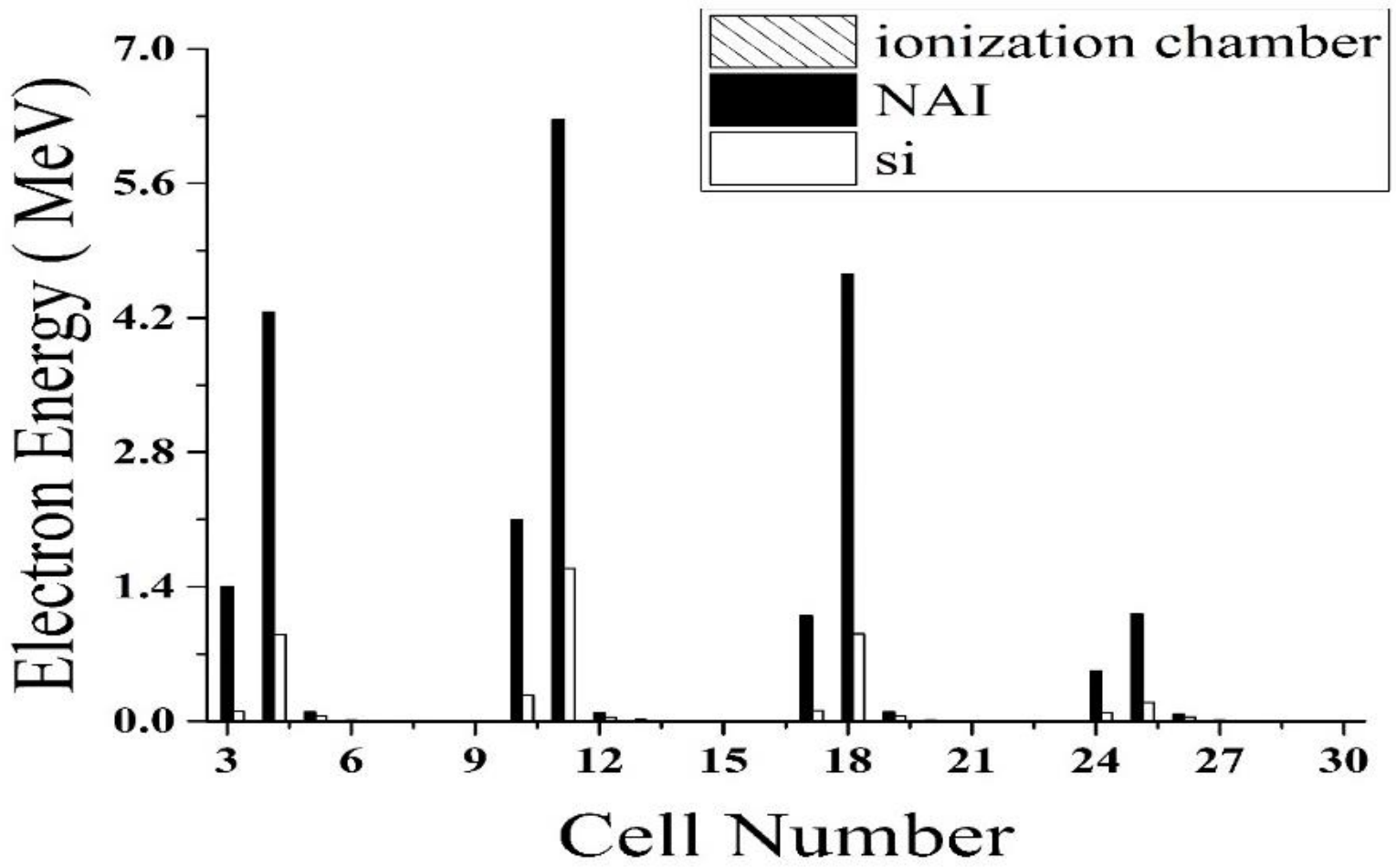

**Figure 11.** Deposited energy by secondary electrons in detectors.

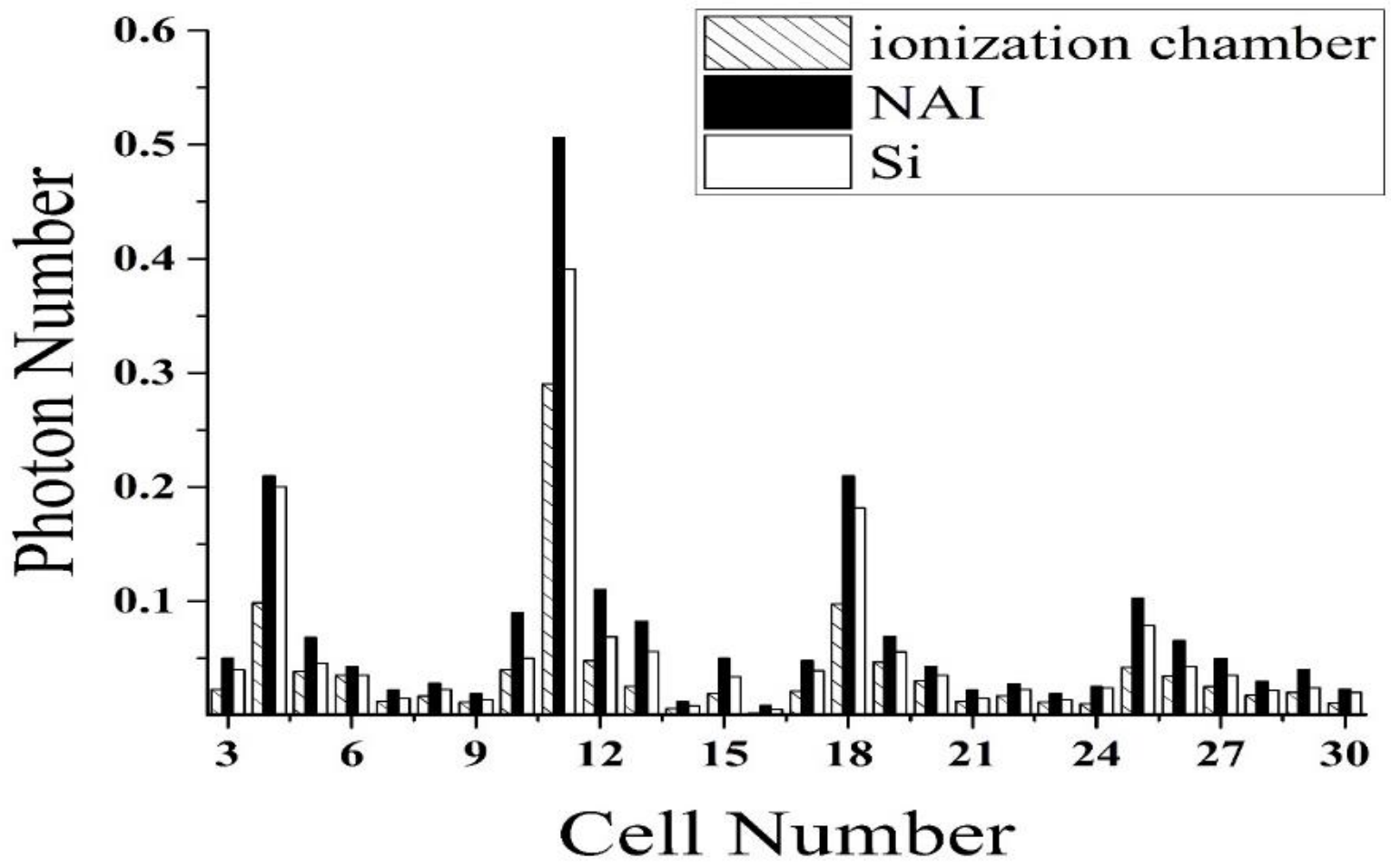

**Figure 12.** The number of deposited photons in detectors.

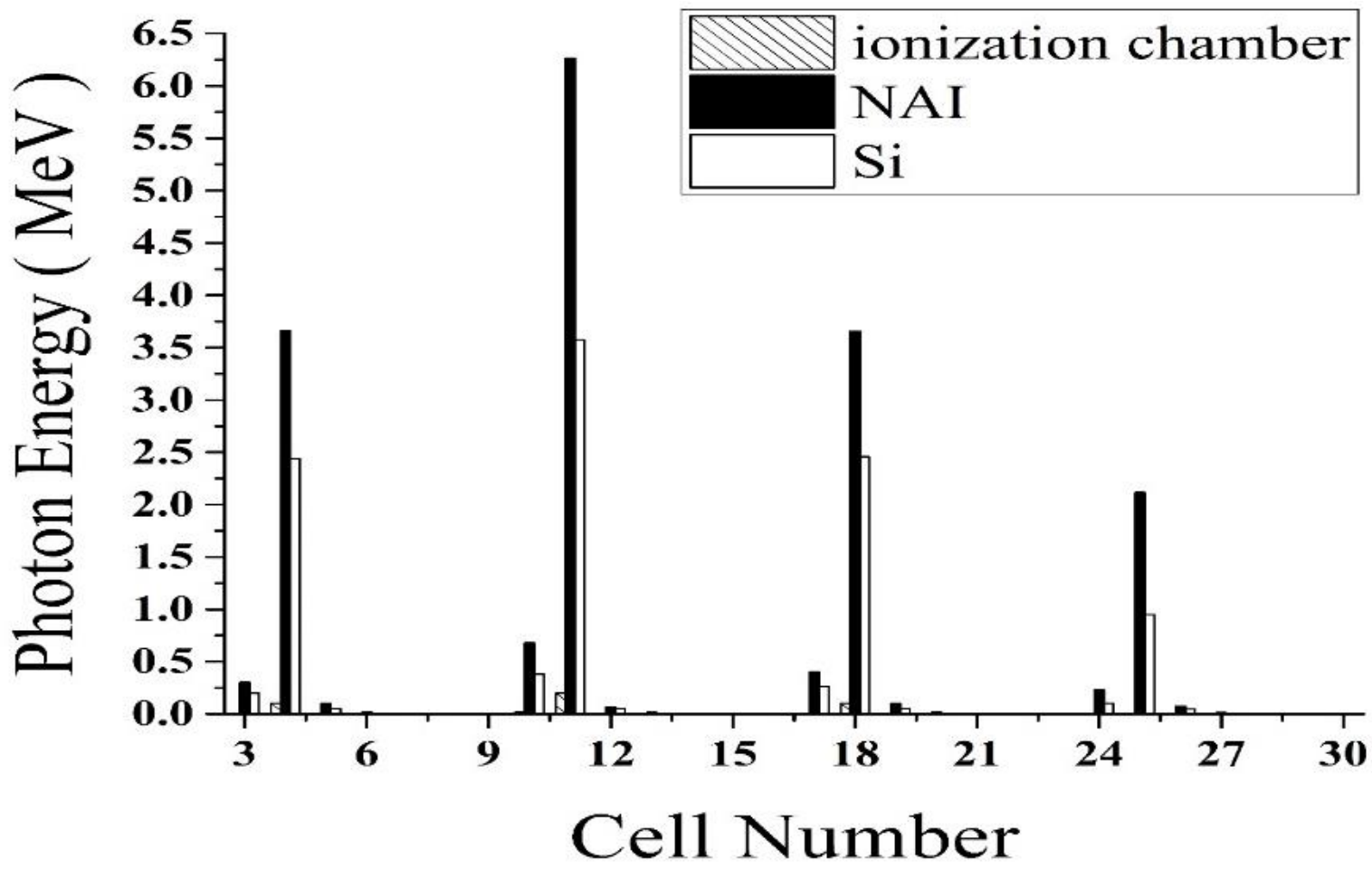

**Figure 13.** Deposited energy by photons in detectors.

As can be interfered from Fig. 10 to 13, most secondary particles reach the cell number 11 and NaI has better results among others in terms of recording the number and deposited energies of secondary electrons and photons. Since for using NaI we need photomultiplier tubes and they are very sensitive to the magnetic fields, therefore, Si detectors can be a good candidate as BLM [32]. The semiconductor silicon detectors have the advantages of high resolution, linear response in a wide range of energies, relatively fast response, high efficiency, ability to work in a vacuum environment, insensitive to the magnetic field, and possibility to be constructed in various forms. Then by placing the Si detector with the same size only in the place of cell 11 and running the simulation again, we will achieve the recorded data as shown in Table V.

**TABLE V.** The number and deposited energy of secondary particles and photons in the Si detector.

| Number of electrons | Deposited energy of electrons (MeV) | Number of photons | Deposited energy of photons (MeV) | Number of neutrons | Deposited energy of neutrons (MeV) |
|---|---|---|---|---|---|
| 0.0543 | 0.1610 | 0.9086 | 3.8981 | 0.5786 | 0.0038 |

## 5. Conclusion

In this paper, we have studied the beam loss due to the magnet misalignment, magnet vibration and magnet current fluctuations that causes the beam to experience a different magnetic field than the nominal value. We have studied the behavior of beam when it sees different magnetic fields inside the magnets. Consequently, we have calculated the deflection angle with respect to the reference beam axis due to the mismatch in the magnetic field of dipole, quadrupole, and sextupole magnets [32]. The deflection angles are obtained to be between 0.5 to 6 degrees for different magnetic fields of dipole magnet and 1.518 to 1.732 degrees for the sextupoles. Therefore, we have considered that electron deflects with the angle of 3 degrees with respect to the beam axis in all further calculations. We have shown that when a 3 GeV electron with 3 degrees of deflection angle hit the beam pipe, it will deposit all its energy there, and finally 34.8 secondary electrons with a total energy of 1073.3020 MeV, 139.7 photons with a total energy of 1467.999 MeV and 0.1 neutrons with a total energy of 1.32 MeV exit from the outer surface of the beam pipe. We have then compared ionization chamber, NaI and Si detectors as candidates for BLM, where it is concluded that NaI has better characteristics for this purpose [3]. However, due to the sensitivity of photomultiplier tubes to the magnetic field, Si detector can be chosen as a second-best candidate in accelerators. Also, it is shown that based on this deflection angle, only some places around the beam pipe are appropriate to place the detector. We have placed the Si detector in 10 cm distance from the beam pipe and have considered the 3 GeV electron to be deflected with the angle of 3 degrees and hit the beam pipe in a simulation done by MCNP [7]. It is concluded that only 0.05 electrons, 0.9 photons and 0.57 neutrons with the total deposited energy of 0.16 MeV, 3.89 MeV and 0.003 MeV, respectively, will reach the detector.